\documentclass{WileyMSP-template}

\usepackage{graphicx}
\usepackage{dcolumn}
\usepackage{bm}
\usepackage{fourier}
\usepackage{comment}
\usepackage{amsmath,manfnt}
\usepackage{siunitx}
\usepackage{soul}
\usepackage[sort&compress,numbers]{natbib}
\usepackage{parskip}
\usepackage{indentfirst}
\usepackage{ragged2e}
\usepackage{float}
\usepackage{xr}
\justifying

\usepackage{totcount}
\newcounter{tofixn}

\regtotcounter{tofixn}
\newcounter{fixedn} 

\regtotcounter{fixedn}

\begin{document}

\pagestyle{fancy}
\rhead{\includegraphics[width=2.5cm]{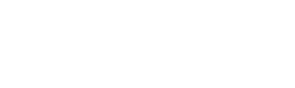}}

\title{Turning Porous Functional Materials into Directional Transport Platforms with Unidirectional Surface Acoustic Waves}

\maketitle


\author{Sujith Jayakumar}
\author{Jinan Parathi}
\author{Gideon Onuh}
\author{Feng Guo}
\author{Ofer Manor}
\author{James Friend*}


\begin{affiliations}
\noindent Sujith Jayakumar, Prof.\ James Friend\\
Medically Advanced Devices Laboratory, Department of Mechanical Engineering and Materials Science, Washington University in St.\ Louis, MO 63130, USA\\
*Email Address: jamesfriend@wustl.edu\\
Jinan Parathi, Gideon Onuh, Prof.\ Ofer Manor\\
Department of Chemical Engineering, Technion – Israel Institute of Technology, Haifa 32000, Israel\\
Prof.\ Feng Guo\\
Intelligent Systems Engineering, Indiana University at Bloomington, IN 47405
\end{affiliations}


\keywords{Acoustofluidics, Porous, Microscale Flows, Acoustic Streaming}


\begin{abstract}
Porous media underpin absorption, filtration, separation, and high-area interfacial transport in chemical and diagnostic systems, yet sustained directional flow through them remains difficult because tortuous pore networks and strong acoustic losses promote bypassing, weak flow, and counterflow. Here, we show that floating-electrode unidirectional transducers (FEUDTs) convert porous materials into actively pumped transport platforms by generating predominantly unidirectional surface acoustic waves (SAWs) that couple more effectively than conventional interdigital transducers across wet multilayer interfaces. By varying pore size, permeability, sample thickness, and fluid viscosity, we find that transport is strongly enhanced when the SAW wavelength is comparable to the characteristic pore dimension, providing a practical design rule for acoustically activated porous media. Under these conditions, FEUDTs drive directional flow velocities up to 0.6 mm s$^{-1}$ at sub-watt input power, about 600 times faster than diffusion alone. FEUDTs also sustain pumping in prewetted porous media, where capillary contributions are removed, yielding velocities that exceed capillary-driven flow under matched conditions while remaining far above thermally induced transport. A reduced theoretical framework captures the main experimental trends and identifies transducer architecture, pore geometry, and actuation strength as the key parameters governing long-range, tunable transport in porous functional materials.
\end{abstract}


\section{Introduction}
\label{sec:introduction}
\par Fluid transport through porous media governs performance in fields ranging from hydrogeology and energy storage to separations, catalysis, and biomedical devices \cite{Bottaro2019Oct, Yi2022Jan}. Their interconnected void network enables the motion of liquids, solutes, gases, and suspended matter, but that same irregular structure also produces tortuous, nonuniform, and sometimes opposing flow paths. As a result, porous materials that are excellent from a materials standpoint may still perform poorly when fluid delivery is rate-limiting. In many systems, the central challenge is therefore not only the chemistry or architecture of the porous medium, but also the ability to drive fluid through it in a controlled and directional manner.

Several approaches have been used to enhance transport in porous structures, including high-pressure pumping \cite{reichmuth2004chip,colon2004very,piendl2021integration}, phase-change or thermally driven methods \cite{qiu2020phase,bachchan2021productivity}, and electric-field-based transport \cite{sprocati2022interplay,lorente2007constructal,revil2007electrokinetic}. Each has important limitations. Pressure-driven systems are often bulky, thermal approaches can be energy intensive, and electrically driven transport commonly depends on fluid polarity or ionic content.

Acoustic actuation is an appealing alternative. Past studies have shown that acoustic waves can transport water through porous media \cite{Li:2007sc,Huang:2020pj,Manor:2021aa,Huang:2022wp}, promote acoustically driven wetting by overcoming contact-line pinning \cite{Li:2007sc}, and improve ion transport through nanoporous battery separators during fast charging \cite{Huang:2022wp}. Because acoustic forcing does not rely on external electric fields or specific fluid chemistry, it offers a compact and versatile route to fluid transport \cite{Huang:2021aa}.

Despite that promise, the mechanism of acoustic transport in fully wetted porous media remains poorly resolved. It is not yet clear whether the dominant wave motion is carried mainly by the liquid phase or by the porous solid matrix, especially in low-loss substrates such as silicon and glass. The coupling among acoustic wavelength, attenuation, pore geometry, and induced flow is also insufficiently understood, particularly for the pore-relevant dimensions of the wavelengths and attenuation lengths associated with MHz-GHz surface acoustic waves (SAWs) \cite{Dentry:2014yk,wu2022manipulations}. Consequently, the roles of acoustic pressure and acoustic streaming remain uncertain in mixed fluid-solid porous systems.

Most SAW-based acoustofluidic devices use an \emph{interdigital transducer} (IDT) as the acoustic source \cite{White:1965aa}. When an IDT is placed directly beneath or within a porous medium, its symmetric generation of SAWs propagating in opposite directions along the substrate in turn generates mutually opposing flow components in the adjacent fluid. This prevents oriented flow in a desired direction. As a consequence, most researchers place the IDT outside the fluid to be actuated, as has been demonstrated particularly well in porous media \cite{Li:2007sc}. Unfortunately, when the IDT is placed outside the fluid-laden porous medium, the propagating SAW is strongly attenuated as soon as it encounters the wetted medium and any intermediate coupling layers. The characteristic leakage length,
$\alpha = \frac{\rho_s c_\text{SAW} \lambda_\text{SAW}}{\rho_f c_f}$,
depends on the fluid and substrate densities, the sound speed in the fluid, the Rayleigh-wave phase speed, and the SAW wavelength \cite{Dentry:2014yk}. The wave therefore leaks rapidly into the fluid and porous matrix \cite{Campbell:1970wb,shiokawa_liquid_1989,vanneste_streaming_2011,Marmottant2017}, confining the transmitted energy to a narrow beam that propagates through the fluid and porous matrix. This significantly limits sustained pumping over distances sufficient for many potential applications in porous media.

Here, that limitation is addressed with a floating-electrode unidirectional transducer (FEUDT), which produces a continuously generated SAW that propagates predominantly in one direction across the full aperture. First introduced by Yamanouchi and Furuyashiki \cite{Yamanouchi1984Nov}, FEUDTs were developed for strongly coupled piezoelectric substrates and later refined extensively for telecommunications applications \cite{Takeuchi2002Aug, Morgan2001Sep, Takeuchi1993Nov, 6725015}. Although the six-finger-per-cell FEUDT geometry is far more demanding to fabricate than a standard IDT because its smallest feature is one-twelfth of the wavelength rather than one-half \cite{Ekstrom2017Feb}, it offers a substantial advantage for porous-media pumping: strong unidirectional SAW generation across the entire active region, which is fortunately much longer than a standard IDT design for the same operating frequency. 

In this work, we establish a route to endow porous functional materials with active, directional transport. We first show that conventional IDTs cannot sustain useful pumping in saturated, acoustically lossy porous media, whereas FEUDTs maintain distributed SAW generation across wet multilayer interfaces and throughout the porous matrix, enabling long-range transport where standard architectures fail. We then show, by systematically varying pore size, sample thickness, and fluid viscosity, that transport is strongly enhanced when the acoustic wavelength is comparable to the characteristic pore dimension, revealing a pore-architecture–wavelength matching rule for converting passive porous substrates into actively pumped transport materials. Finally, we demonstrate that this concept extends to a biologically relevant dermis-like matrix, in which FEUDT actuation drives rapid, directional small-molecule transport. Together with quantitative analysis of viscous and acoustothermal effects and a reduced theoretical framework that captures the principal experimental trends, these results establish practical design rules for acoustically activated porous materials and position acoustically activated porous media as a versatile materials platform for transport-limited chemical, diagnostic, and biomedical systems.

\section{Results and Discussion}
\label{sec:results_discussion}
\subsection{Waveform generated on an IDT vs FEUDT}
\label{subsec:ldv_scans}
\begin{figure*}
\centering
\includegraphics[width=0.55\linewidth]{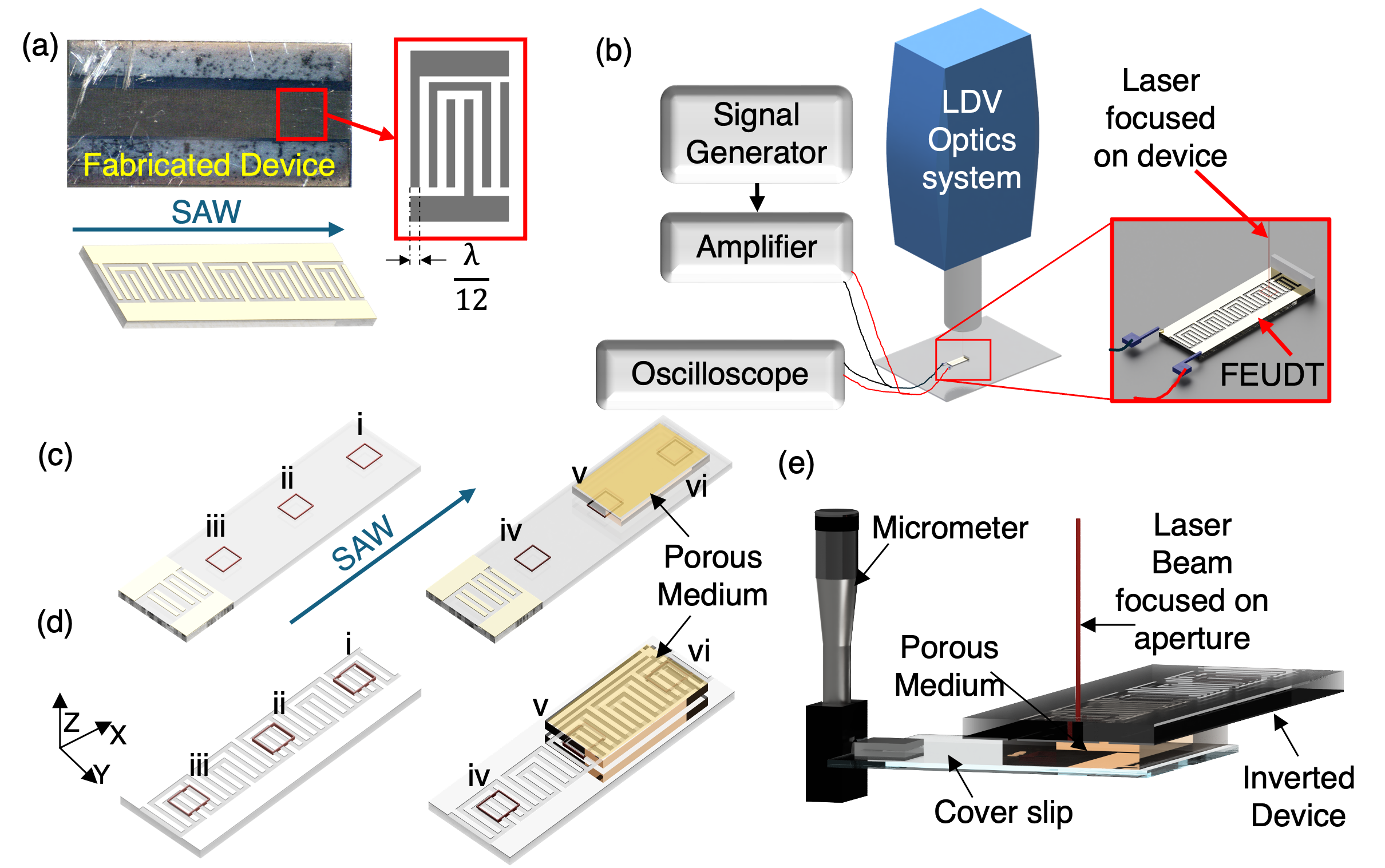}
\caption{\label{fig:ldv_setup} (a) Fabricated FEUDT device and its single unit cell (red box). (b) Laser Doppler vibrometer setup to scan the waveform generated by the bare device. (c) Regions (i), (ii), and (iii) denote scan regions on a bare lithium niobate substrate for an IDT. Regions (iv), (v), and (vi) correspond to scans on the substrate before the porous media, at its entrance, and further upstream within the porous media, respectively. (d) Equivalent scan regions for the FEUDT. (e) Inverted FEUDT scanning configuration, in which the laser passes through the 0.5~mm thick substrate to focus on the aperture in contact with the porous media, held in place by a micrometer-controlled glass cover slip.}
\end{figure*}

The IDT and FEUDT were compared by measuring the out-of-plane SAW amplitude along the $x$-direction under identical operating conditions, with particular attention to attenuation within a fluid-laden porous medium. A laser Doppler vibrometer (LDV; UHF-120SV, Polytec, Irvine, CA, USA) was used to record the vibration amplitude at multiple positions (Figure~\ref{fig:ldv_setup}(a,b)). Each transducer was driven with a sinusoidal electrical signal across its resonance window (38--42~MHz) to identify its operating frequency. Point-by-point LDV scans were then performed at resonance, within $\pm 0.01$~MHz, on a predefined grid with a spatial resolution of $\lambda/8$, where the designed SAW wavelength was $\lambda = c_\text{s}/f_\text{op} = (3840~\mathrm{m/s})(40\times10^6~\mathrm{Hz})^{-1} = 96~\mu\mathrm{m}$. Here, $c_\text{s}$ is the sound speed in the substrate and $f_\text{op}$ is the operating frequency. To improve measurement fidelity, each point was defined by complex averaging over eight acquisitions. All measurements were performed under identical excitation and experimental conditions to enable direct comparison.

Both the IDT and FEUDT generated nearly ideal unidirectional traveling SAWs under the measurement conditions. Upon scanning the lithium niobate surface of the transducer, a largely unattenuated SAW traveling in the positive $x$-direction with nearly uniform amplitude was observed: for the IDT, anywhere in front of the aperture (Figure~\ref{fig:ldv_setup}(c)); and for the FEUDT, throughout the aperture (Figure~\ref{fig:ldv_setup}(d)), as indicated by positions (i), (ii), and (iii) in Figure~\ref{fig:ldv_setup}(c,d). The measurement uncertainty for both configurations was $\pm 0.68$~mm/s. The IDT measurements were obtained at an acoustic particle velocity of $7.58$~mm/s, a frequency of $39.8$~MHz, and $V_\text{RMS}=2.51$~V, whereas the FEUDT measurements were obtained at $7.39$~mm/s, $40.58$~MHz, and $V_\text{RMS}=2.51$~V. The standing wave ratio (SWR) for both devices was calculated as $\mathrm{SWR}=\max{|d|}/\min{|d|}$, where $d$ is the displacement amplitude measured by LDV. The resulting SWR values were 1.031 for the IDT and 1.033 for the FEUDT at positions (i), (ii), and (iii) in Figure~\ref{fig:ldv_setup}(c,d), confirming that both devices produced nearly pure traveling waves in this configuration.

Although both the IDT and FEUDT generated nearly ideal traveling SAWs on the bare lithium niobate substrate, only the FEUDT sustained SAW propagation once the porous medium was placed under wet loading. This distinction shows that the key advantage of the FEUDT is not simply unidirectionality on an unloaded substrate, but the ability to continue generating SAW beneath the loaded region itself. To test this directly, the devices were mounted in the inverted configuration shown in Figure~\ref{fig:ldv_setup}(e), such that the porous medium contacted the substrate in front of the IDT (Figure~\ref{fig:ldv_setup}(c)) and across the FEUDT aperture (Figure~\ref{fig:ldv_setup}(d)). A coverslip mounted on a micrometer stage provided precise and repeatable contact, and the LDV laser was directed through the LN substrate to measure the out-of-plane displacement at the substrate-porous-medium interface. On the bare substrate immediately before the porous medium (position (iv) in Figure~\ref{fig:ldv_setup}(c,d)), both devices produced waveforms that closely matched those measured at position (i), confirming comparable traveling-wave behavior before fluid loading. At the entrance to the porous medium (position (v)), a wave traveling in the positive $x$-direction was still detected for both devices, with SWR values of 1.663 for the IDT and 1.002 for the FEUDT. The decisive difference appeared farther into the medium at position (vi): for the IDT, the SAW was strongly attenuated at the porous-medium interface and no detectable propagation remained along the $x$-direction, whereas the FEUDT sustained propagation beneath the wet porous layer with an SWR of 1.110, close to its unloaded value of 1.033. We attribute this difference primarily to the distributed placement of FEUDT electrodes throughout the loaded region, which permits continued local SAW generation despite mass loading by the fluid-saturated porous structure. In contrast, the IDT launches the wave upstream of the load, so the wave must survive attenuation at the interface without further reinforcement. These results show that distributed generation under load, rather than unidirectionality alone, is what enables FEUDTs to maintain long-range coupling and drive fluid transport in wet porous media.

\subsection{Flows driven by an IDT and its limitations}
\label{subsec:whatman_idt}
\begin{figure*}
\centering
\includegraphics[width=0.95\linewidth]{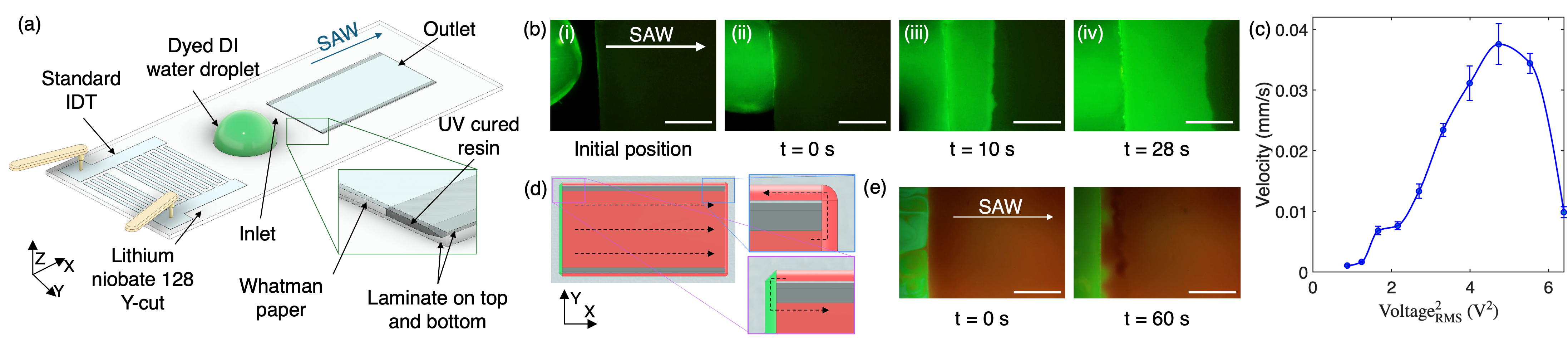}
\caption{\label{fig:whatman_idt} (a) Schematic of the Whatman paper sample coupled with the IDT via a 10~$\mu$m-thick UV epoxy layer. Inset: porous media with sealed edges (using UV resin) and laminated surfaces to prevent flow along the sides or beneath the sample. (b) Meniscus movement of green dye in a dry Whatman paper under IDT actuation at $V_\text{RMS} = 5.07$~V. (c) Fluid velocities (blue line) plotted against $V_\text{RMS}$, after subtracting capillary-driven contributions (0.081~mm/s), showing the SAW-induced component. Error bars indicate the 95~\% confidence interval (CI), calculated from velocities independently obtained across six experiments (n=6) by fitting the slope of each distance–time profile. (d) No bulk fluid flow is observed in prewetted porous media under IDT actuation. (e) Limited penetration of the green-dyed fluid in a prewetted porous medium (dyed red) under SAW actuation. Scale bar: 1~mm.}
\end{figure*}

Laminated Whatman paper samples were used to evaluate the coupling efficiency of surface acoustic waves generated by conventional IDTs with porous media. Specifically, laminated AE100 Whatman paper (GE Healthcare, Chicago, IL, USA) samples measuring 4~mm $\times$ 6~mm were prepared for flow characterization. The rationale for lamination is described in Section~\ref{subsec:laminate_reason}. Each sample was bonded directly to the substrate in front of the standard IDT aperture using a 10~$\mu$m-thick layer of UV-curable epoxy resin (Norland Optical Adhesive 81, Edmund Optics, Barrington, NJ, USA); Figure~\ref{fig:whatman_idt}(a). The inlet and outlet were aligned with the SAW propagation axis ($x$-axis) so that fluid transport occurred along the same direction.

In an initially dry porous sample, SAW actuation increased fluid transport beyond the capillary-driven baseline, although the measured meniscus velocity reflected contributions from both mechanisms. A 10~$\mu$L droplet of deionized (DI) water mixed with fluorescent dye was pipetted 0.5~mm upstream of the inlet and, upon actuation, was driven through the porous sample until it reached the outlet. The advancing meniscus was tracked from sequentially captured images (Figure~\ref{fig:whatman_idt}(b)), yielding a velocity of 0.104~mm/s at $V_\text{RMS}=2.17$~V. This measured velocity includes the capillary-driven contribution. The capillary-driven flow alone produced an initial (maximum) fluid velocity of 0.04~mm/s under an estimated capillary pressure of 11.25~kPa, $\sim \left(2\sigma \cos\theta / r\right)$, where $\sigma=72.8~$mN/m is the liquid-air surface tension, $\theta=22^\circ$ is the equilibrium contact angle, and $r=12~\mu$m is the characteristic pore radius of the medium. By contrast, SAW actuation further increased the meniscus velocity because acoustic radiation pressure at the fluid-air interface within the porous medium drove the meniscus through the pore network \cite{Bok:2009kx}. Acoustic streaming, which scales as $\rho U^2$, generates pressures on the order of only 100~Pa, indicating that streaming alone is insufficient to drive fluid across an air-water interface. The corresponding SAW-induced enhancement, obtained after subtracting the capillary contribution, is plotted against $V_\text{RMS}^2$ in Figure~\ref{fig:whatman_idt}(c). The data exhibit an approximately linear relationship ($R^2 \approx 0.97$) up to $V_\text{RMS}=2.17$~V, consistent with prior studies \cite{Friend2011Jun}. At higher input voltages, however, the velocity decreased, primarily because droplet atomization and jetting reduced the fluid volume available for pumping.

IDT actuation did not produce meaningful long-range flow through the fully wetted porous medium, because the acoustic energy was dissipated near the inlet. To isolate the IDT-driven contribution and eliminate capillary-driven flow, the sample was first prewetted with water (dyed red; 10~\% v/v dye in DI water; Red Water tracing dye, EcoClean, Copiague, NY, USA). To preserve fluid mass continuity, the inlet and outlet were connected by a thin fluid return passage in the laminated sample (Figure~\ref{fig:whatman_idt}(d)), creating a racetrack-like configuration. The porous medium was prewetted with the red dye, and a green-dyed tracking fluid (10~\% v/v dye in DI water; Green Water tracing dye, EcoClean, Copiague, NY, USA) was introduced to visualize transport. Under actuation, the green dye was expected to traverse the porous medium and emerge at the outlet. Instead, it penetrated less than 0.2~mm from the inlet even at $V_\text{RMS}=5.5$~V (Figure~\ref{fig:whatman_idt}(e)). This limited propagation is attributed to rapid SAW attenuation at the droplet interface, which dissipates the acoustic energy locally. As a result, acoustic streaming remains confined near the sample entrance and generates insufficient pressure to drive flow through the wetted porous medium.

\subsection{Flows driven by an FEUDT}
\label{subsec:whatman_feudt}
\begin{figure*}
\centering
\includegraphics[width=0.75\linewidth]{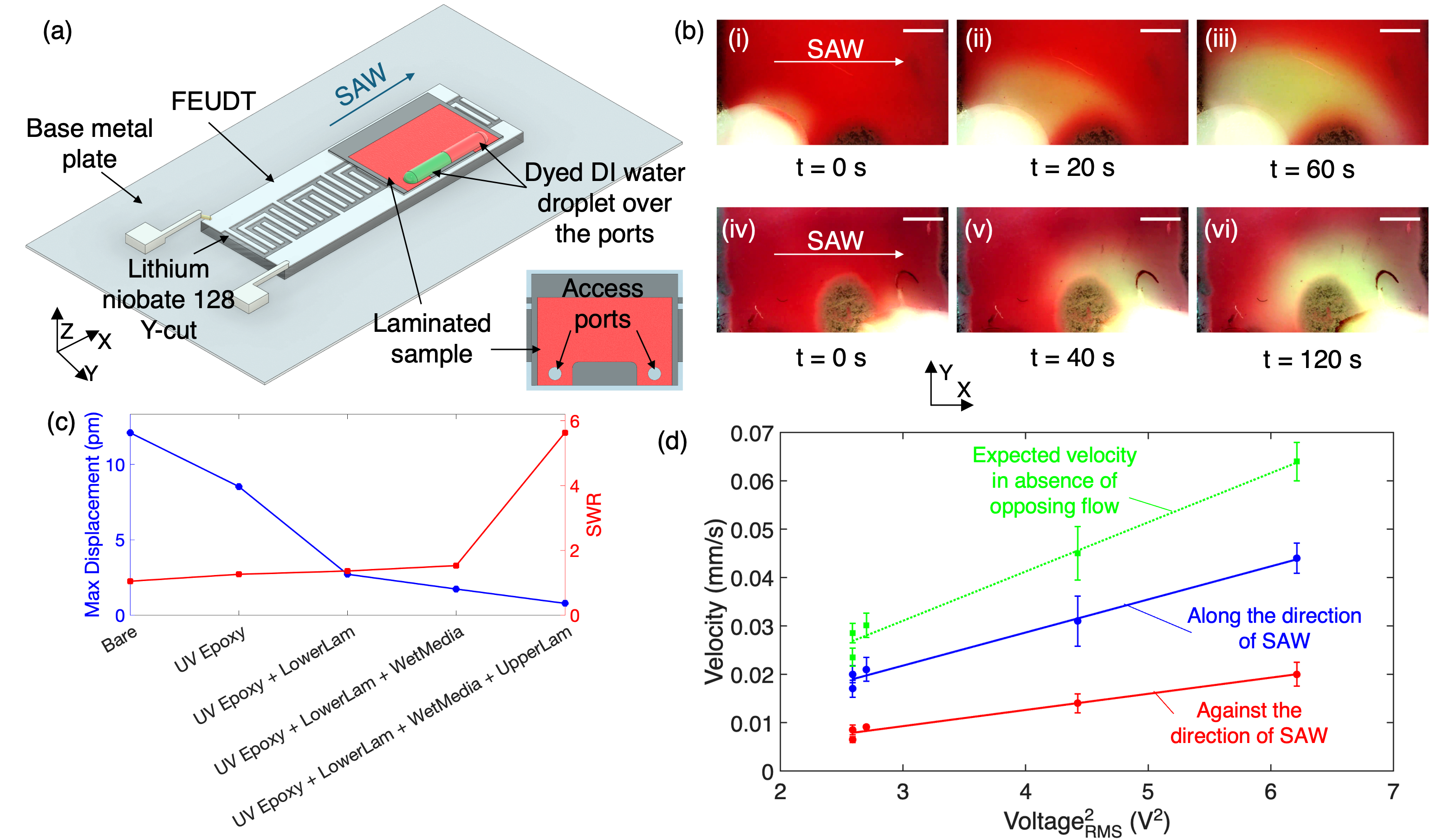}
\caption{\label{fig:whatman_feudt} (a) Schematic of the laminated Whatman paper sample coupled with the FEUDT, showing access ports and droplets positioned to complete the closed fluid circuit. (b) Pumping behavior at $V_\text{RMS} = 2.49$~V, illustrating the movement of green dye through the prewetted (red-dyed) porous medium (i–iii) along the SAW propagation direction and (iv–vi) against it. Scale bar: 1~mm. (c) Attenuation of the FEUDT-generated SAW across successive coupling layers, showing the increase in standing wave ratio (SWR, red line) and the concurrent decrease in displacement amplitude (blue line) by the time the wave reaches the porous medium. (d) Flow velocities measured for configuration (a) at different input voltages, demonstrating faster flow in the SAW direction (blue) compared to the opposing direction (red); the anticipated ideal velocity, in the absence of reverse flow components, is indicated in green. Error bars: 95~\% CI; n=6.}
\end{figure*}

FEUDT actuation enabled substantially faster long-range SAW-driven flow through the \textit{wet} porous sample than was achieved with conventional IDTs (Figure~\ref{fig:whatman_feudt}). In this configuration, the green dye was expected to traverse the porous medium and emerge at the outlet containing the red droplet, thereby circulating through the medium. Meniscus tracking showed flow velocities reaching 0.044~mm/s at $V_\text{RMS}=2.49$~V (Figure~\ref{fig:whatman_feudt}(b)(i--iii)). Diffusion was negligible under these conditions, because the diffusion time scale, $T_\text{D}=L^2/D \approx 5\times10^3$~s, was roughly three orders of magnitude larger than the acoustic-streaming time scale, $\mathcal{O}(\beta U)^{-1}\sim 5.22$~s \cite{Friend2011Jun}. Here, $D=5\times10^{-3}~\mathrm{mm}^2/\mathrm{s}$ and $L=5$~mm, giving a diffusion-only velocity of $D/L=0.001$~mm/s.

\subsection{The attenuating effect of coupling layers and pore size}
As intermediate coupling layers were added between the FEUDT and the porous medium, the wave amplitude progressively decreased and the waveform departed from a purely traveling SAW, indicating increasing attenuation and partial reflections within the multilayer structure. To examine this coupling behavior across all intermediate layers, LDV scans were conducted after sequentially introducing the UV epoxy, lower laminate, Whatman paper, and top laminate. The maximum displacement magnitudes and standing wave ratio (SWR) values for each configuration are shown in Figure~\ref{fig:whatman_feudt}(c). The displacement amplitude decreased by approximately one order of magnitude, from 12.1~pm on the bare substrate to 1.73~pm over the wet Whatman paper. At the same time, the Supporting Information shows the gradual emergence of partial standing-wave characteristics. Consistent with this trend, the SWR increased from 1.046 on the bare substrate to 1.529 on the wet Whatman paper, confirming increasing reflection within the multilayer stack. Flow measurements further supported this interpretation. When the positions of the green and red dye droplets were reversed, fluid motion was also observed against the SAW propagation direction (Figure~\ref{fig:whatman_feudt}(b)(iv--vi)), although at less than half the velocity of the forward flow (Figure~\ref{fig:whatman_feudt}(d)). These results suggest that, although the net flow remains in the positive $x$-direction, weak opposing components arise from reflected waves; reducing these reflections could further enhance the forward flow velocity, as suggested by the green curve in Figure~\ref{fig:whatman_feudt}(d).

Although FEUDT actuation enabled rapid pumping even in pores as small as 12~$\mu$m, the attainable flow rate was likely limited by the wavelength-pore-size mismatch and by attenuation across the intermediate coupling layers. Each FEUDT unit cell has a width equal to the SAW wavelength on the substrate ($\lambda = 96~\mu$m), and the repeated unit cells generate SAWs that constructively interfere across the aperture to produce a unidirectional wave field. In the present system, however, the characteristic pore size of the porous medium ($\sim 12~\mu$m) is substantially smaller than $\lambda$, which may hinder efficient coupling between the SAW and the porous structure. This mismatch can promote partial reflections and localized caustics, potentially generating flow components in random or opposing directions, consistent with earlier observations showing flows in opposition to the nominal SAW propagation direction. In addition, reducing the number of coupling layers is expected to decrease wave attenuation, allowing a larger fraction of the SAW amplitude to contribute directly to fluid pumping. These considerations motivated the efforts in the following section to further improve the flow rate.

\subsection{Increasing the pore size to approximate the SAW wavelength in the substrate}
\label{subsec:pe_results}
\begin{figure*}
\centering
\includegraphics[width=0.95\linewidth]{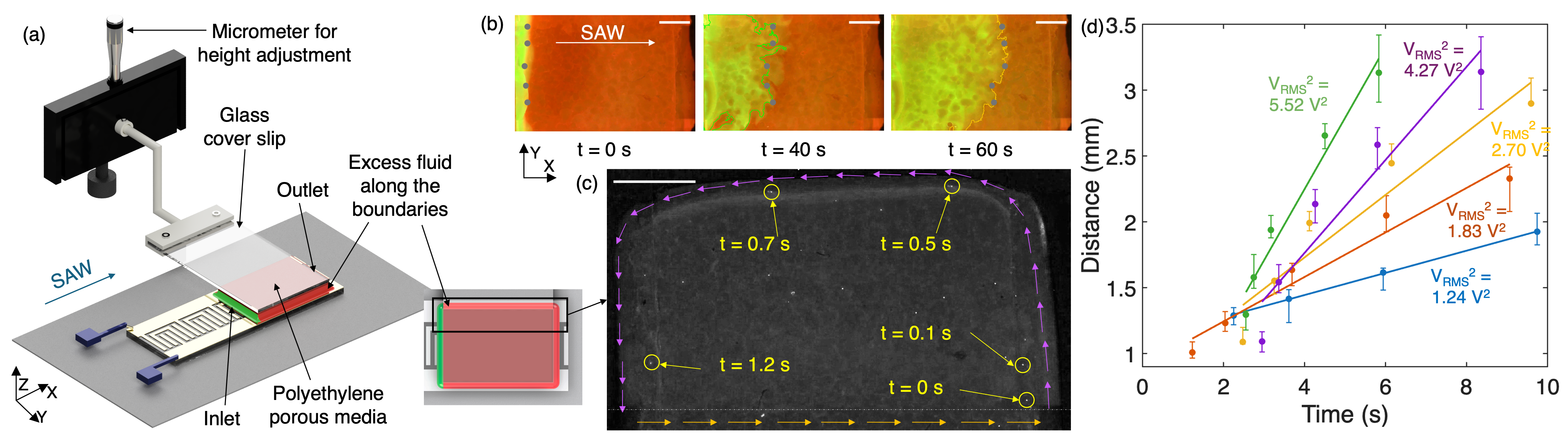}
\caption{\label{fig:PE_results} (a) Schematic of the FEUDT coupled with a polyethylene (PE) porous medium, illustrating the coverslip used to adjust the height and maintain conformal contact between the PE sample and the FEUDT surface, thereby preventing flow above or below the sample. (b) Pumping of green dye through the prewetted (red-dyed) porous medium at $V_\text{RMS} = 1.12$~V. (c) A series of images over time that indicate recirculating flow with 10.2~$\mu$m fluorescent particles. Magenta arrows indicate flow in the external boundary fluid, while orange arrows denote flow through the porous medium induced by the FEUDT. (d) Distance traveled by the dye within the porous medium as a function of time for different applied voltages. Data points represent the mean distance from six independent ($n=6$) experiments. For each experiment and each time, the distance was computed as the average of five predefined points (grey points in (b)) sampled along the advancing fluid meniscus at fixed y-axis (vertical) positions across the aperture. Error bars denote the minimum and maximum values across the six trials. Solid lines represent linear fits to the data. The coefficients of determination ($R^2$) were consistently high across all datasets, with a minimum value of 0.90. Scale bar: 1~mm.}
\end{figure*}

We replaced the Whatman paper with a commercially available polyethylene (PE) porous media whose pore size more closely matched the SAW wavelength in the substrate to approach the ideal case in which the FEUDT more effectively drives flow through the porous medium by acting on the fluid menisci within the pore network. We anticipate that matching the pore size to the SAW wavelength (i.e., the FEUDT cell width) enhances coupling by increasing the effective interaction area, enabling more efficient transfer of acoustic energy and directionality to the fluid within the pores. In contrast, smaller pores may promote increased acoustic reflections and caustic formation, limiting distributed coupling and favoring more localized, less directional fluid transport. The PE medium (Porex Corporation, Fairburn, GA, USA) had a polydisperse pore-size distribution of 60--100~$\mu$m (porosity: 94\%; permeability: $1.84\times10^{-11}~\mathrm{m}^2$), which approximates the substrate SAW wavelength of 96~$\mu$m. The sample was prewetted with red-dyed DI water and placed directly on the FEUDT, with surrounding boundary fluid as in the previous experiments (Figure~\ref{fig:PE_results}(a)). A green-dyed DI water droplet was then introduced near the entrance on the left side of the sample. A glass coverslip was positioned above the porous medium to ensure full contact with the FEUDT while suppressing external flow. To avoid applying excess pressure, the coverslip was held by an external mount attached to a micrometer stage with 1~$\mu$m precision, allowing controlled height adjustment.

The PE porous medium supported complete fluid circulation under FEUDT actuation at relatively low input voltage, and dye proved more effective than particles for estimating the flow velocity. Upon actuation, the green-dyed fluid was pumped through the red-dyed region at a velocity of 0.05~mm/s at $V_\text{RMS}=1.12$~V (Figure~\ref{fig:PE_results}(b)) and exited through the outlet on the right. To verify recirculation around the porous sample, 10.2~$\mu$m fluorescent polystyrene (PS) particles suspended in DI water were used as tracers. Near the outlet, the particles were observed to loop around the porous sample outside the FEUDT-electrode region and return to the inlet, thereby confirming the recirculating flow path shown in Figure~\ref{fig:PE_results}(c). Dye, however, was more suitable for estimating meniscus velocity, because the irregular pore geometry trapped particles and complicated single-particle tracking in the three-dimensional flow.

The measured transport is consistent with Darcy flow driven by SAW-induced acoustic streaming, with a steady velocity that scales linearly with $V_\text{RMS}^2$. Once wetted, the porous medium supports flow through pressure gradients generated at the pore scale by SAW-induced Reynolds stresses in the fluid \cite{NYBORG1965265}. These pressure gradients drive bulk transport according to Darcy's law. In SAW-driven systems, both the pressure gradient and its source term, the first-order acoustic particle velocity, scale linearly with the applied voltage, since $u_1 \propto \omega A \propto V_\text{RMS}$, where $A$ is the SAW amplitude. The resulting acoustic streaming velocity is therefore a second-order nonlinear effect, giving $u_\text{streaming} \propto u_1^2 \propto V_\text{RMS}^2$ \cite{Friend2011Jun}. Consistent with this picture, Figure~\ref{fig:PE_results}(d) shows that, at fixed actuation, the dye displacement increases linearly with time, $x(t)=U_\text{streaming}t$, indicating transport at an approximately constant velocity. The slope of each distance-time curve gives the steady streaming velocity, which in turn increases linearly with $V_\text{RMS}^2$ (Figure~\ref{fig:pore_height_variation}), as expected for pressure-gradient-driven pumping. Although the larger-pore PE medium also has a higher intrinsic permeability and therefore would be expected to support greater Darcy flow, the enhancement observed here likely reflects the combined effects of permeability and improved acoustic coupling; the latter interpretation is supported by the directional-flow comparison and by the thickness dependence discussed below.

\subsection{Comparing IDT and FEUDT-driven flows for different pore sizes and sample heights}
\label{subsec:pore_height_variation}
\begin{figure*}
\centering
\includegraphics[width=0.5\linewidth]{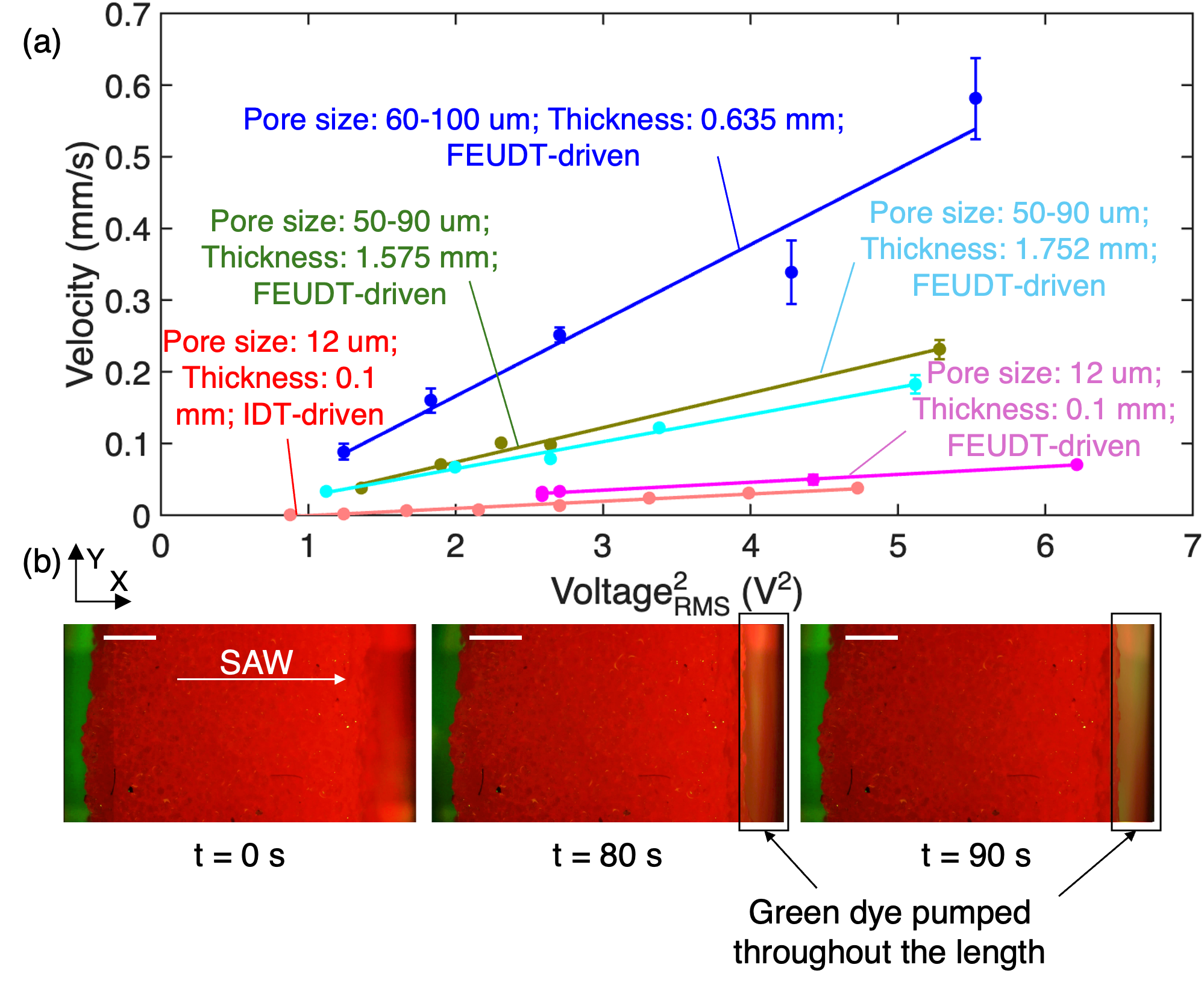}
\caption{\label{fig:pore_height_variation} (a) Pumping velocity as a function of pore size and sample thickness. Enhanced flow is observed when the pore size is comparable to the SAW wavelength in the substrate and when the observation plane is closer (thinner samples) to the FEUDT surface. Error bars indicate the 95~\% confidence interval, calculated from velocities independently obtained across six experiments (n=6) by fitting the slope of each distance–time profile. (b) For a thicker sample (H = 3.125~mm), no visible green dye or pumping behavior is observed on the top surface, although traces of dye at the outlet end (black boxes) confirm complete pumping through the sample. Scale bar: 1~mm.}
\end{figure*}

The IDT and FEUDT were compared under matched acoustic excitation, allowing differences in flow behavior to be solely attributed to the distribution of the acoustic waves as generated by the two SAW transducers. Although electrical voltage was the controlled input, the physically relevant actuation parameter for acoustic streaming is the acoustic particle velocity at the substrate surface. To establish this equivalence, LDV measurements of particle velocity were performed as a function of input voltage for both transducer types under identical loading conditions, with the wet PE sample placed on the device. As shown in Figure~1 in of the Supporting Information, similar voltage actuation ranges produced comparable particle-velocity amplitudes for the IDT and FEUDT, thereby justifying the use of equal voltage inputs for direct comparison. The particle velocity showed a slightly less than linear dependence on voltage, scaling approximately as $V_\text{RMS}^{0.92}$ for the IDT and $V_\text{RMS}^{0.82}$ for the FEUDT. We speculate that the departure from ideal linear scaling in these relationships are due to mass loading and damping introduced by the porous medium, which was intentionally included during these measurements to reproduce the conditions under which SAW-driven flow occurs.

FEUDTs produced faster and more directional flow than IDTs, and flow was further enhanced when the characteristic pore size more closely matched the acoustic wavelength, even in relatively thick porous media. For pores as small as $12~\mu$m, FEUDT-driven systems outperformed IDT-driven systems regardless of any capillary contribution. Within the FEUDT-based configurations, comparison of the $12~\mu$m Whatman paper with the $60$--$100~\mu$m polyethylene (PE) porous media showed substantially faster directional flow in the latter, despite the observation plane being much higher above the substrate (0.635~mm versus 0.1~mm), as shown in Figure~\ref{fig:pore_height_variation}(a). Although larger pores inherently permit higher flow rates, these results also indicate that closer matching between the acoustic wavelength and the characteristic pore dimensions enhances directional pumping by reducing acoustic caustics and reflections that can generate opposing flow components, as observed earlier. To assess SAW propagation in thicker samples, PE porous media with 50--90~$\mu$m pores were examined at two heights, 1.57~mm and 1.75~mm, while keeping the pore size distribution fixed. At the same input $V_\text{RMS}$, the green dye reached the upper observation plane later in the thicker sample, indicating a lower flow speed. These results show that FEUDTs can nevertheless drive flow through relatively thick porous media. Although higher velocities are expected closer to the FEUDT surface because of acoustic attenuation with height, only the upper plane could be visualized experimentally. For a 3.125~mm-thick sample, the dye traversed the full sample length, as shown in Figure~\ref{fig:pore_height_variation}(b), without appearing in the upper plane, confirming that SAW-induced flow persists throughout the porous medium even at substantial thickness.

\subsection{The effects of changing the viscosity, including thermal effects}
\label{subsec:plots_viscous_effects}
\begin{figure*}
\centering
\includegraphics[width=0.95\linewidth]{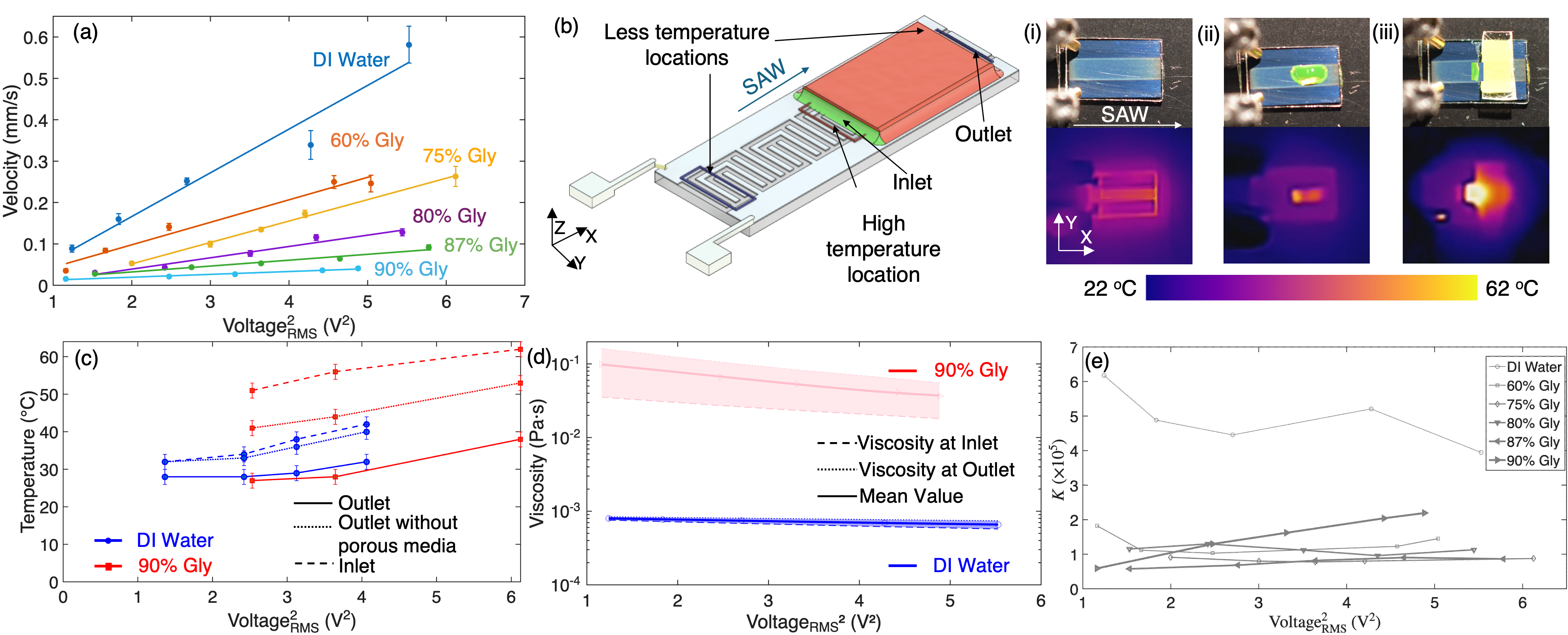}
\caption{\label{fig:plots} (a) Variation of flow velocity for fluids with different viscosity, as indicated in the plot. Error bars: 95~\% CI; $n=6$. (b) Schematic representation showing the measurement locations of maximum and minimum temperatures when the FEUDT was actuated under three configurations: (i) bare substrate, (ii) with a droplet on the aperture, and (iii) with a polyethylene porous medium covered by a glass coverslip. These images were captured using a thermal IR camera. (c) Fluid temperatures measured using a contact thermocouple at the locations indicated in (b) for DI water (blue) and 90~\% glycerol (red). Error bars: min–max; n=5. The temperatures correspond to values recorded 0.5~s after actuation. (d) Mean viscosity (solid lines) and viscosity difference across the sample ends, estimated from the temperature variation for DI water (blue) and 90~\% glycerol (red). (e) Nondimensional number signifying the ratio of acoustic stresses over viscous stresses for different water-glycerol mixtures at varying voltage inputs.}
\end{figure*}

The measured flow did not decrease in proportion to the nominal fluid viscosity, contrary to the simplest expectation for acoustic streaming. To examine this behavior, water-glycerol mixtures spanning a wide viscosity range were tested, and the resulting flow velocities were plotted against $V_\text{RMS}^2$ (Figure~\ref{fig:plots}(a)). Although increasing viscosity would ordinarily be expected to suppress acoustic streaming, the 90~\% glycerol solution, whose viscosity is approximately 175 times that of DI water at 25~$^{\circ}$C, did not exhibit the proportionally reduced flow rate predicted from nominal viscosity alone.

To determine whether heating itself could account for the observed transport, we measured the fluid temperature near the inlet and outlet of the FEUDT, both with and without the porous medium (Figure~\ref{fig:plots}(b)). On the bare substrate, actuation at $V_\text{RMS}=2.20$~V produced a nearly uniform temperature rise to approximately 35~$^{\circ}$C across the aperture, which we attribute to Joule heating arising from Ohmic losses in the FEUDT electrodes (Figure~\ref{fig:plots}(c)). When either a 90~\% glycerol droplet or a porous medium prewetted with the same fluid and covered with a glass slip was placed on half of the device, the maximum temperature in both cases increased to 62~$^{\circ}$C at the same location, indicating acoustothermal heating caused by viscous damping in the fluid \cite{Kondoh2005Oct, Kondoh2009Feb, Das2025Jun}. The identical peak temperatures in these two configurations suggest that viscoelastic attenuation within the solid matrix is negligible. The inlet region exhibited the highest temperature, likely because constructive SAW generation by upstream electrodes enhanced local viscous dissipation, whereas the downstream region was driven primarily by the electrodes directly beneath the fluid. The outlet temperature was higher for the droplet than for the porous medium, suggesting that the porous structure suppresses convective redistribution of heat. Importantly, however, even a maximum temperature difference of 25~$^{\circ}$C (Figure~\ref{fig:plots}(c)) yields a characteristic thermally induced flow of only 12.3~nm/s, estimated from $v_\text{char}=J/\rho_\text{avg}$ with $J=D\rho_\text{grad}$, which is approximately $10^{4}$ times smaller than the measured SAW-driven flow. Heating therefore does not directly drive the observed transport.

Instead, the principal role of heating is to modify the local viscosity of the working fluid, particularly for the more viscous glycerol-water mixtures. Over the short transport length considered here (half the device length, 5~mm), the viscosity variation generated by the inlet-to-outlet temperature gradient can be approximated as linear. Temperature-dependent properties of the water-glycerol mixtures were estimated following prior reports \cite{Cheng2008May,Volk2018Apr}, and the corresponding mean viscosity and inlet-to-outlet viscosity difference are shown in Figure~\ref{fig:plots}. The effective viscosity decreased linearly with increasing $V_\text{RMS}^2$, implying that the instantaneous flow velocity was highest near the inlet, where the fluid was warmest and least viscous, and then decreased approximately linearly along the sample as the temperature fell and the viscosity increased, as illustrated in Supplementary Information Figure~2. This spatial variation is consistent with a local linear relationship between viscosity and pumping velocity. The observed linear increase in streaming velocity with $V_\text{RMS}^2$ ($R^2 = 0.982$; Figure~\ref{fig:plots}(a)) is therefore consistent with SAW-driven pumping operating in a fluid whose effective viscosity decreases during actuation. In this sense, acoustothermal heating does not generate the transport, but it does explain why the measured velocity departs from scaling based on the nominal fluid viscosity alone.

The transport nevertheless remains in a viscous streaming regime across the full actuation range examined. To assess this, we compared the acoustically generated stresses with the viscous stresses by defining the dimensionless stress ratio $K=\sigma_\text{acoustic}/\sigma_\text{viscous}=\rho u_{\mathrm{ac}}^2/(\mu v_s/\ell)$, where $\rho$ and $\mu$ are the fluid density and dynamic viscosity, $u_{\mathrm{ac}}$ is the acoustic particle velocity, $v_s$ is the measured steady-state streaming velocity, and $\ell$ is a characteristic pore length scale. Figure~\ref{fig:plots}(e) shows that $K$ depends only weakly on $V_{\text{RMS}}^2$ across all water-glycerol mixtures, indicating that increasing the electrical input scales acoustic forcing and viscous dissipation nearly in proportion. This behavior is consistent with a steady-streaming regime governed by a balance between acoustic Reynolds stresses and viscous dissipation associated with attenuation of the acoustic wave \cite{Friend2011Jun}. The glycerol-rich fluids collapse into a relatively narrow range of $K$, whereas DI water exhibits systematically higher values, suggesting that the more viscous mixtures remain firmly in a viscous-dominated streaming regime while DI water lies somewhat closer to a transitional regime in which viscous dissipation is less dominant.

\subsection{Theory}

\begin{figure}[htbp]
    \centering
    \includegraphics[width=0.5\textwidth]{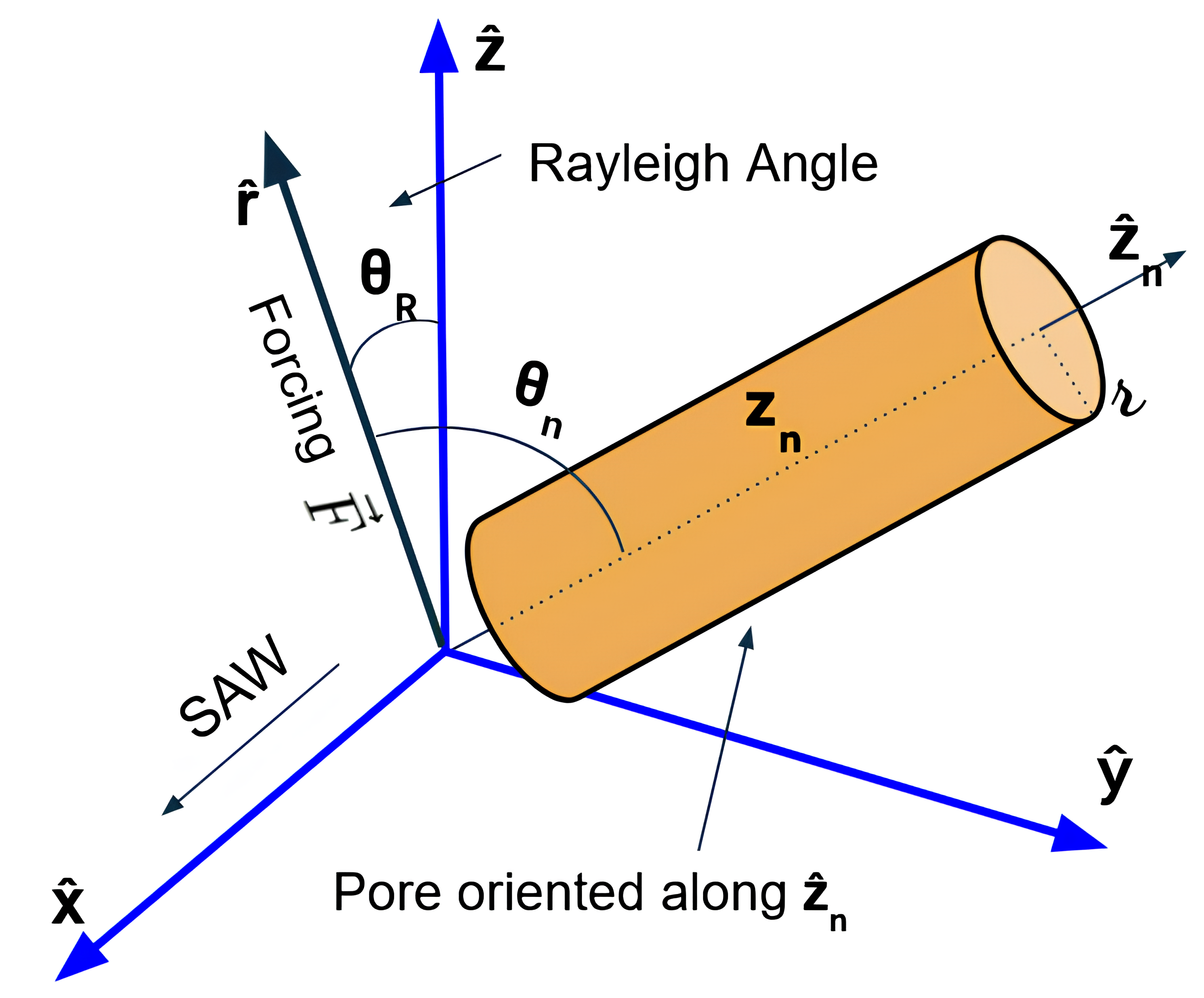}
    \caption{Coordinate system and geometric configuration of the model. The surface acoustic wave (SAW) propagates along $\hat{x}$, inducing a forcing vector $\vec{F}$ at the Rayleigh angle $\theta_R$. A representative pore is oriented along the unit vector $\hat{z}_n$ at an angle $\theta_n$ relative to the forcing direction.}
    \label{fig:coordinate_diag}
\end{figure}

To interpret the experimental results, we introduce a reduced model for SAW-induced acoustic streaming in a liquid-saturated porous medium. The aim is not to resolve the full elastic response of a heterogeneous porous solid, but to capture the leading dependencies of the flow on actuation amplitude, attenuation, pore size, and sample thickness. We therefore consider an idealized half-space of porous medium in contact with a FEUDT that supports a unidirectional Rayleigh SAW. The SAW propagates along the solid substrate and leaks longitudinal ultrasound of the same frequency into the porous medium at the Rayleigh angle, $\theta_R$. The coordinate system and pore geometry are shown in Figure~\ref{fig:coordinate_diag}.

Because the porous materials used here are highly porous and consist of cellulose- or elastomer-based matrices with densities comparable in magnitude to those of the liquid phase, we approximate the saturated porous region as an effective medium with a coarse-grained acoustic impedance similar to that of water at the ultrasound wavelength scale. Under this approximation, the particle-velocity components of the leaked ultrasound parallel and normal to the FEUDT surface, along $x$ and $z$, are given by \cite{Fasano2025}
\begin{equation} \label{eq:Ux_complex}
    U_{x} = \frac{i k_{s} U}{\sqrt{k_{s}^2 - k^2}} \exp\left(-z \sqrt{k_{s}^2 - k^2} - i k_{s} x + i \omega t\right), \text{and}
\end{equation}
\begin{equation} \label{eq:Uz_complex}
    U_{z} = U \exp\left(-z \sqrt{k_{s}^2 - k^2} - i k_{s} x + i \omega t\right),
\end{equation}
where $U$ is the normal amplitude of the SAW particle velocity at the FEUDT surface, $k_s$ and $\omega$ are the SAW wavenumber and angular frequency, respectively, and $t$, $x$, and $z$ denote time and the coordinates parallel and normal to the FEUDT surface. The ultrasound wavenumber in the porous medium is $k \equiv \omega/c - i\alpha$, where $c$ is the coarse-grained ultrasound phase velocity in the porous medium and $\alpha$ is the corresponding attenuation coefficient.

Dissipation of this acoustic field generates a time-averaged streaming force, $\mathbf{F}$, through the divergence of the Reynolds stress,
\begin{equation}\label{eq:force_exp}
    \mathbf{F} = -\rho \langle (\mathcal{R}(\mathbf{U}) \cdot \nabla)\mathcal{R}(\mathbf{U}) + \mathcal{R}(\mathbf{U}) \nabla \cdot \mathcal{R}(\mathbf{U}) \rangle,
\end{equation}
where $\rho$ is the liquid density, $\mathcal{R}\left(\mathbf{U}\right)$ denotes the real part of the velocity vector $\mathbf{U}=\left(U_{x},~U_{z}\right)$, and
\[
\langle \chi \rangle\equiv\lim_{T\to\infty}\frac{1}{T}\int_{t=0}^{t\to T}\chi \,\mathrm{d}t
\]
denotes time averaging over long times compared with $\omega^{-1}$. To leading order, the force along the ultrasound propagation path at the Rayleigh angle, $\theta_R = \cos^{-1}(F_s^{-1})$, scales with the square of the SAW velocity amplitude:
\begin{equation}\label{eq:force_RayleighAngle}
    F_r = F_{S} C_{i} U^2 \rho \exp(-2 C_{i} z),
\end{equation}
where $F_s \equiv \sqrt{ \left(\left( {k_s C_r}\right)/\left({C_i^2 + C_r^2} \right)\right)^2 + 1 }$ is a dimensionless factor, and $C_r$ and $C_i$ are the real and imaginary parts of $\sqrt{k^2 - k_s^2}$, respectively.

We next model the porous medium as a network of $N$ randomly oriented cylindrical pores of characteristic diameter $R_p$. Flow along each pore is driven by the streaming force in Equation~\eqref{eq:force_RayleighAngle} and opposed by viscous dissipation. To represent the radial shape of the pore-scale velocity profile, we introduce a curvature parameter $b$. For fully developed Poiseuille flow in a smooth cylindrical pore, $b=1/R_p$. As inertial flattening or interfacial slip becomes more important, the profile becomes less curved and $b$ decreases, approaching $b=0$ in the ideal plug-flow limit.

In the present low-Reynolds-number setting, it is useful to interpret this reduced curvature through an effective slip length at the pore wall. Surface roughness in combination with viscous dissipation can generate an apparent slip length \cite{10.1063/1.4982899}, $\beta\approx\epsilon_1\epsilon_2\mu/\sqrt{\rho p_\circ}$, which increases with the liquid shear viscosity $\mu$, where $\epsilon_1$, $\epsilon_2$, and $p_\circ$ denote the ratio of roughness scale to $r_p$, a geometric factor, and the characteristic pressure drop along the pore, respectively. For Poiseuille flow in a cylindrical pore with Navier slip length $\beta$, the radial curvature can be written as $b=1/\sqrt{\beta r_p}$. We therefore define
$b \equiv \frac{1}{n r_p}$, so that $n=\frac{1}{b r_p}=\sqrt{\beta/r_p}$. The dimensionless factor $n$ thus serves as an effective measure of pore-scale slip relative to pore size.

Averaging the mass transport along the SAW propagation direction over random pore orientations and over the acoustic forcing across a porous layer of thickness $h$ gives the effective bulk velocity along the substrate,
\begin{equation}\label{eq:final_average_velocity}
    u = \frac{\zeta (nr_{p})^2 F_{S} C_{i} U^2 \rho}{4\mu} \left[1 - \frac{1}{2n^2} \right] \frac{\sin(\theta_{R})}{3} \left[ \frac{1 - e^{-2 C_{i} h}}{2 C_{i} h} \right].
\end{equation}
Here, $\zeta$ is the porosity. Equation~\eqref{eq:final_average_velocity} shows that the average streaming velocity scales as $U^2$, consistent with classical Eckart streaming, and is modulated by attenuation through $C_i$ and by pore morphology through $r_p$, $h$, and $n$. A full derivation is provided in the Supporting Information.

Laser Doppler vibrometer measurements show that the SAW normal velocity amplitude on the FEUDT scales linearly with the applied root mean square voltage. Using the least-squares relation $U = aV$, with $a\approx 4.2$ volt$\cdot$s/mm, Equation~\eqref{eq:final_average_velocity} becomes
\begin{equation}\label{eq:final_power_velocity}
    u=\frac{\zeta r_{p}^2 F_{S} C_{i} a^2 V^2 \rho}{4\mu} \left(n^2 - \frac{1}{2} \right) \frac{\sin(\theta_{R})}{3} \left( \frac{1 - e^{-2 C_{i} h}}{2 C_{i} h} \right).
\end{equation}
This form directly predicts the experimentally observed quadratic scaling, $u \propto V^{2}$, while also retaining the effects of viscosity, attenuation, pore size, and sample thickness.

To compare the model with experiment, temperature-dependent viscosity changes were included using the measured inlet and outlet temperatures for each condition. Because acoustothermal heating produces a spatial viscosity gradient along the porous sample, the corresponding flow velocities were evaluated using the average temperature for each case. For each porous medium, the characteristic pore size used in the model was taken as the median pore diameter of that material.

\begin{figure}[H]
    \centering
    \includegraphics[width=0.6\textwidth]{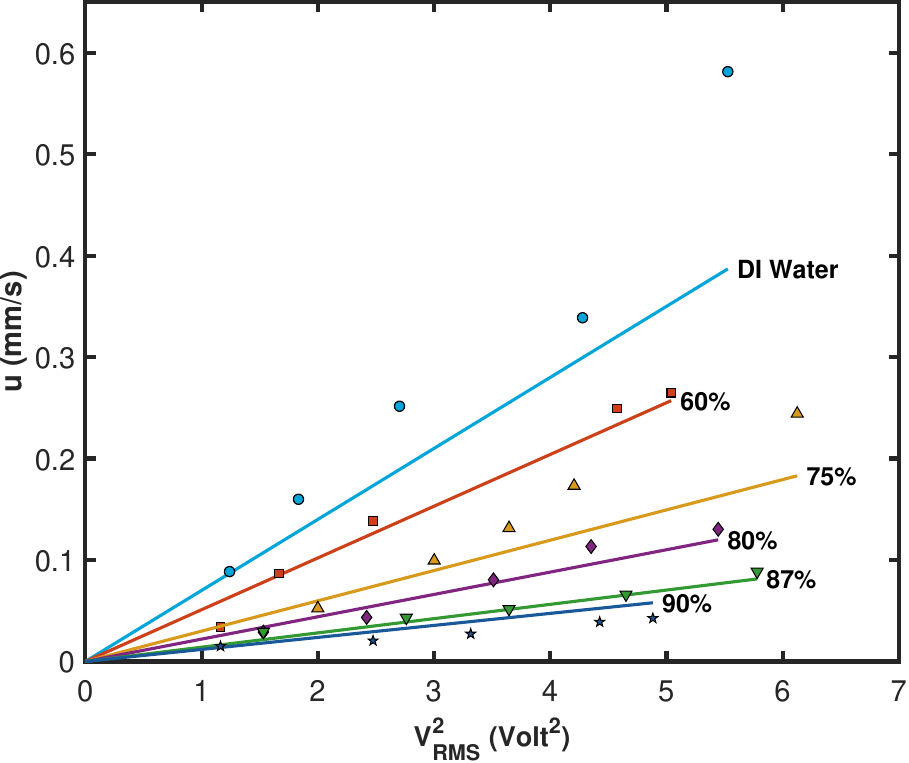}
    \caption{Voltage squared ($V_{\mathrm{RMS}}^{2}$) dependence of the measured (symbols, also shown in Figure~\ref{fig:plots}(a)) and simulated (solid lines) flow velocities through porous polyethylene for different glycerol concentrations. For the highly viscous water/glycerol mixtures, $n = 30$ is used, whereas $n = 21$ is used for DI water. The viscosity values correspond to a liquid temperature of $T = 30^{\circ}\mathrm{C}$.}
    \label{fig:theory1}
\end{figure}

\begin{figure}[H]
    \centering
    \includegraphics[width=0.6\textwidth]{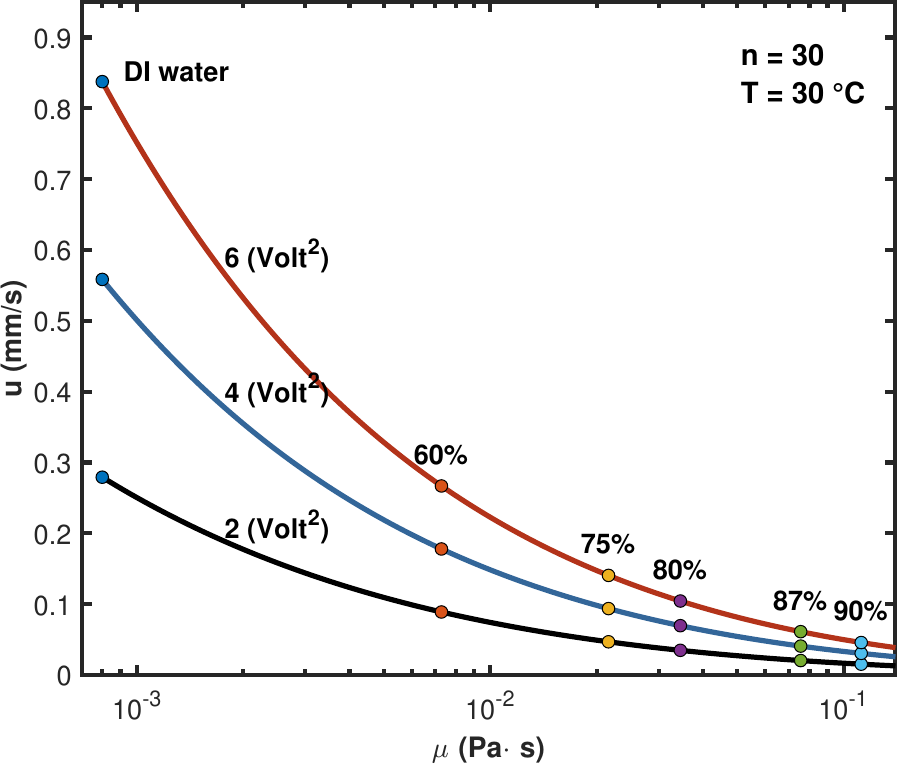}
    \caption{Flow velocity, $u$, in porous polyethylene as a function of shear viscosity, $\mu$, for different applied voltage-squared values, $V^2$, using the same framework as in Figure~\ref{fig:theory1}. The solid curves represent the theory for $n = 30$, and the symbols correspond to the working fluids (DI water and glycerol--water mixtures of varying concentration) positioned according to their viscosity values at $T = 30^{\circ}\mathrm{C}$.}
    \label{fig:theory2}
\end{figure}

\begin{figure}[H]
    \centering
    \includegraphics[width=0.6\textwidth]{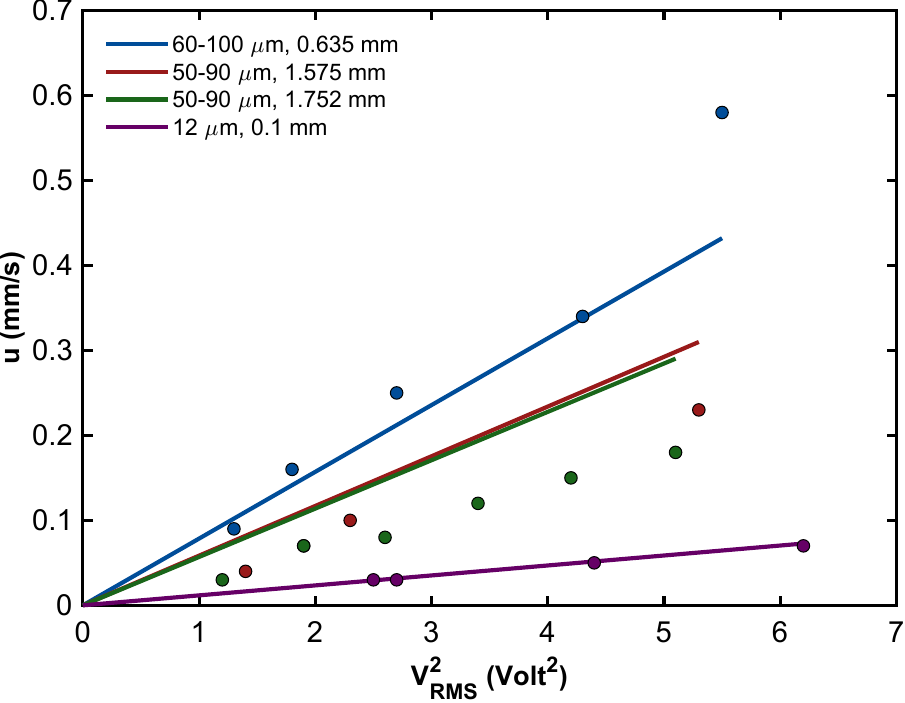}
    \caption{Voltage squared ($V_{\mathrm{RMS}}^{2}$) dependence of the measured (symbols, also shown in Figure~\ref{fig:pore_height_variation}(a)) and simulated (solid lines) DI water flow velocities, $u$, through the different porous media used in the experiments. The calculations use the viscosity of water at $T = 30^{\circ}\mathrm{C}$ and $n=21$.}
    \label{fig:theory3}
\end{figure}

The comparison between theory and experiment supports a second-order acoustic-streaming response, in which the steady pore-scale transport scales with the square of the ultrasound particle velocity $u \propto V^{2}$, as in classical Eckart streaming \citep{eckart_vortices_1948}. For water over the range of excitation used here, the corresponding acoustic Reynolds number, $\rho U \lambda/\mu \approx 0.1$, remains below unity, and is smaller still for the more viscous glycerol--water mixtures, where $\lambda$ is the ultrasound wavelength.

A related Reynolds number based on pore size, $\rho U r_p/\mu \approx 0.1$, also remains below unity for the porous systems considered here. In porous polyethylene, where the average pore size is comparable to the ultrasound wavelength, the pore-scale Reynolds number and the acoustic Reynolds number are of similar magnitude. In the Whatman paper, by contrast, the characteristic pore size is much smaller, approximately one quarter of the wavelength, so the pore-scale Reynolds number is correspondingly lower. These small Reynolds numbers justify the viscous-flow framework adopted here, while still allowing for a weak convective correction through the parameter $n$. Additional details are given in the Supporting Information.

Figure~\ref{fig:theory1} compares the measured liquid velocity in porous polyethylene with the theoretical prediction. Good qualitative agreement is obtained using $n=21$ for DI water and $n=30$ for the more viscous glycerol--water mixtures. Within the present framework, the increase in $n$ with viscosity is consistent with an increased effective slip length at the pore surface, arising from the coupling among roughness, viscous stresses, and the local pressure field. Thus, the lower-viscosity DI water corresponds to a smaller effective slip parameter, whereas the more viscous glycerol--water mixtures are better described by larger values of $n$.

To isolate the effect of viscosity from that of the effective slip length, Figure~\ref{fig:theory2} holds $n$ constant and examines the predicted variation of flow velocity with viscosity for different excitation levels. This representation highlights the strong sensitivity of the flow to both viscosity and applied voltage, and shows that the experimentally explored range spans nearly an order of magnitude in transport velocity.

Figure~\ref{fig:theory3} further compares the model with the DI water experiments performed in different porous media. Reasonable agreement is obtained using the same value, $n=21$, across these configurations. Because surface roughness, pore geometry, and pore-size distribution differ among these materials, this result suggests that, within the present set of porous media, the influence of liquid viscosity on the effective slip parameter is more important than the variations associated with pore morphology alone.

\subsection{Demonstration of the utility of the FEUDT for enhancement of drug delivery through subcutaneous layer of porcine skin}
\label{subsec:skin_tests}
\begin{figure*}
\centering
\includegraphics[width=0.99\linewidth]{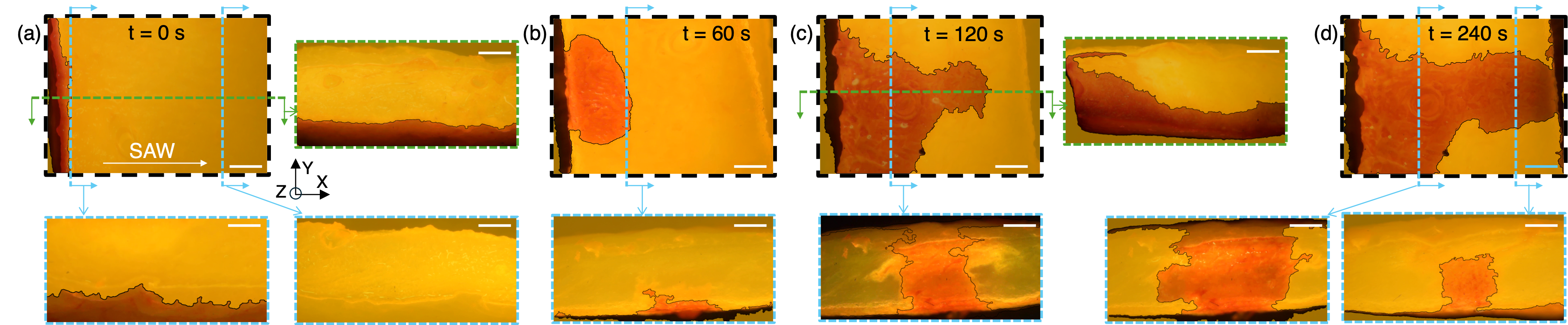}
\caption{\label{fig:skin_tests} Yellow regions correspond to porcine skin, while red regions indicate rhodamine B transported due to acoustic actuation. Images with black-dashed borders show planar views of the dermal layer that was in contact with the FEUDT aperture and subsequently reversed at the indicated time points to capture the pumped dye profile. Blue and green dashed lines mark the cross-sections along which the skin was dissected to visualize dye penetration and transport. (a) At $t = 0$~s, no rhodamine B penetration is observed in the upstream cross-sections. (b) After 60~s of actuation, limited penetration in the thickness direction ($z$) is observed, while transport predominantly occurs along the positive $x$-direction. (c) At 120~s, increased penetration depth (in $z$-direction) is evident in the same cross-section, accompanied by further downstream transport ($x$-direction); side cross-sections (green) reveal enhanced penetration (in $z$-direction) near the inlet that increases with time relative to the corresponding location in (a). (d) By 240~s, rhodamine B has traversed the full length of the sample and penetrated the entire thickness (in $z$-direction) near the inlet. Scale bar: 1~mm.}
\end{figure*}

FEUDT-generated surface acoustic waves (SAWs) markedly enhanced transport of the small-molecule tracer Rhodamine B (MW $\approx 479$~Da) through porcine dermis, but not through intact skin. Porcine skin is a well-established surrogate for human soft tissue because of its similar collagen density, extracellular matrix (ECM) architecture, and permeability \cite{Ranamukhaarachchi2016Aug}. When intact skin, including the stratum corneum, epidermis, and dermis, was positioned with the stratum corneum facing the FEUDT aperture, no dye permeation was observed. This result is consistent with the minuscule nanometer-scale aqueous pathways of the stratum corneum \cite{Itoh2008Jun} that prevent flow. It also indicates that the applied acoustic energy does not disrupt the epidermal barrier and that FEUDT-generated SAW-driven acoustic streaming requires micrometer-scale porosity comparable to the relevant wavelength scale, as suggested by the results in the previous sections.

Exposing the dermal layer enabled robust unidirectional acoustically driven transport that greatly exceeded passive diffusion. When the sample was reversed so that the dermis faced the FEUDT, its high porosity (82--90~\%) and pore diameters of 80--100~$\mu$m \cite{Vallecillo2021Aug, Huang2022Aug} supported efficient SAW-driven flow. At 0.77~W, the Rhodamine B front advanced 5.9~mm along the $x$-axis within 240~s, corresponding to a mean velocity of 0.024~mm/s, and traversed the full 1.5~mm tissue thickness within approximately 120~s. These values are consistent with prior SAW microfluidic studies and demonstrate that acoustic streaming can overcome diffusional bottlenecks in compliant biological matrices \cite{Wang2025Apr}, as long as the stratum corneum is breached. Notably, while acoustically enhanced transdermal transport has been widely explored, the present FEUDT-based approach enables controlled, long-range, and directionally guided flow through the tissue, distinguishing it from conventional BAW-based methods \cite{Li2025Dec, Zhu2024May} that primarily demonstrate penetration without comparable spatial control. Using reported diffusion coefficients for Rhodamine-type dyes, $D \approx 3.1$--$3.6\times10^{-8}$~cm$^{2}$/s \cite{Wu2021Feb}, a purely diffusing molecule would travel only $L=\sqrt{2Dt}\approx 0.041$~mm in 240~s, far less than the acoustically driven distance of 5.9~mm. Likewise, complete penetration through 1.5~mm of tissue within 120~s exceeds the corresponding diffusion-limited distance of 0.029~mm by more than an order of magnitude.

Crucially, the transport enhancement observed in porcine dermis is relevant beyond this model system because Rhodamine B is a well-established surrogate for sub-500~Da therapeutics, and similar extracellular transport barriers arise in many biological matrices. Rhodamine B is widely used as a proxy for small-molecule drugs because of its similar molecular weight, lipophilicity, and transport pathways \cite{Xu2023Feb, Forster2012Mar}. Fluorescence-based dye infiltration in porcine skin is also a validated predictor of small-molecule mobility, supporting its use here as a mechanistic analogue \cite{Kennedy2017Nov}. Because the dermal extracellular matrix of porcine skin resembles that of subcutaneous connective tissue \cite{Jacobi2007Feb}, tumor stroma \cite{Netti2000May}, fibrotic organs, and hydrogel-based organ models \cite{Wolf2012Oct}, these findings may extend to a broad range of translational settings. Small-molecule agents of similar size, including leuprolide acetate (469~Da), buserelin (484~Da), and triamcinolone (434~Da), could therefore benefit from acoustically enhanced penetration \cite{Diao2013Oct}. Potential applications include improving drug dispersal from subcutaneous depots, increasing chemotherapeutic infiltration into stromal tissue, and overcoming transport resistance in fibrotic organs \cite{Stylianopoulos2012Sep}. In organ-on-chip systems, SAWs could likewise provide directional and controllable solute delivery through collagen, GelMA, or fibrin matrices \cite{Bhatia2014Aug}.

Collectively, these findings show that FEUDT-generated SAWs enable selective transport through permeable extracellular matrix (ECM) networks, provide directional flow that can be tuned by acoustic power, and produce transport rates far exceeding diffusion. The transport distance can, in principle, be extended by increasing the FEUDT aperture length, allowing the pumping range to be tailored to the application. These characteristics position SAW actuation as a promising strategy for on-demand subcutaneous and interstitial drug delivery in tissues with dermis-like microstructure, and potentially in settings where the stratum corneum is breached via microneedles or similar methods.

\section{Experimental Section}
\label{sec:experimental_materials_methods}
A design frequency of $40$~MHz was selected, as this value avoids spurious bulk wave generation in the 500~$\mu$m thick lithium niobate substrate and ensures proper SAW confinement \cite{Hashimoto2004}. The piezoelectric substrate used was a $500$~$\mu$m thick single-side polished $127.68$ degree Y-cut X-propagating lithium niobate (Roditi International Corporation, London, England). The IDT and FEUDT devices were fabricated by the photolithography process on a 1~$\mu$m negative photoresist (NR9-1500PY, Futurrex, Franklin, NJ, USA) exposed to a laser writer (Heidelberg Instruments MLA150, Heidelberg, Germany) at a dose of $1000$~mJ/cm$^2$. After developing using the RD6 developer, $20$~nm chromium followed by $350$~nm aluminum using sputtering using a Denton Discovery $18$ (Denton Vacuum, Inc., Moorestown, NJ, USA). The IDT was fabricated with 24 electrode pairs and 12 reflectors on one side to generate a unidirectional SAW on the substrate. Each electrode finger was designed to be one-quarter of the SAW wavelength in the substrate ($\approx25~\mu$m), corresponding to an actuation frequency of 40~MHz. For the same operating frequency, the FEUDT employed finger widths equal to one-twelfth of the SAW wavelength ($\approx8.25~\mu$m). Each FEUDT comprised a total of 100 unit cells (Figure~\ref{fig:ldv_setup}(a)). The number of finger pairs or unit cells in both devices was selected to obtain a designed impedance of $50~\Omega$, thereby ensuring proper matching with the RF driving circuit. The waveform generated by FEUDT was tested using a laser Doppler vibrometer scan (Polytec UHF-120SV, Polytec, Inc., Irvine, CA, USA) by connecting to an RF amplifier (Amplifier Research 10U100, Microwave Instrumentation, Souderton, PA, USA), for which the input is given by a signal generator WF1968 (NF Corporation, Yokohama, Japan). A UV fluorescent dye (Water tracing dye, EcoClean, Copiague, NY, USA) and fluorescent particles of diameter $10.2$~$\mu$m (Fluoresbrite YG Microspheres, Polysciences, Warrington, PA, USA) were used as a tracer inside the fluid to observe the fluid motion. For visualization, they are subjected to a $450$~nm $80$~mW blue laser (Edmund Optics Inc, Barrington, NJ, USA).  

The samples of porous media used in the study include Whatman paper (AE100, GE Healthcare, Chicago, IL, USA) and polyethylene porous sheets (Porex Corporation, Fairburn, GA, USA). A glass coverslip (AmScope, Irvine, CA, USA) was used to control the height of the porous sample using a differential micrometer drive (DRV304, Thorlabs Inc, Newton, NJ, USA). Porcine skin sheets (Stellen Medical LLC, Fisher Scientific, Hampton, NH, USA) were obtained in a shaved and frozen state, with a uniform thickness of $1.524$~mm $\pm$ $0.254$~mm. Rhodamine B (R6626, St. Louis, MO, USA) served as the tracer for experiments involving porcine skin. We used a digital microscope (407-Pro, Andonstar, Shenzhen, China) to visualize and record the fluid motion within the porous samples. The resulting image sequences were processed in a Python environment (code provided in the supplementary information) to track the meniscus and extract flow velocities. The algorithm identified the advancing front by detecting color transitions, using the intensity shift between the green dye and the prewetted red substrate to determine the boundary position at each time point. For every frame, the boundary location was measured at five positions across the FEUDT aperture and averaged to obtain a representative wavefront position. This procedure was repeated over multiple time instants, and the resulting averaged distance–time data were linearly fitted to compute flow velocities. Six independent experiments were performed for each configuration and the final velocities were calculated from the mean distance propagated across all replicates. A high-speed camera (Fastcam Mini UX100, Photron, Tokyo, Japan) is used with a long-distance microscope (Infinity K2, Infinity Photo-Optical Company, Centennial, CO, USA) lens with an objective lens (K2 Close-Focus Objective CF-4) for monitoring the fluid height between porous sample and the transducer. The temperatures were recorded using a thermocouple (HH911T, Omega, Michigan City, IN, USA) and a thermal imaging camera (HP96, HSF Tools, Vietnam). 

\subsection{Rationale for Laminating the Whatman Paper Sample}
\label{subsec:laminate_reason}
To visualize and confine flow within the porous sample, two UV-fluorescent dyes (red and green) were used to distinctly observe fluid motion. A 4~mm~($x$) × 4~mm~($y$) piece of porous media was positioned in front of the IDT and coupled in a wet state to ensure effective SAW transmission. However, it was observed that the acoustically driven flow predominantly occurred above and below the porous sample rather than within it. This occurred because the pores presented a higher hydraulic resistance compared to the thin fluid films that formed around the sample, causing the Whatman paper to partially float on a fluid layer. When the dispensed fluid volume exceeded the saturation capacity of the porous medium, this thin interfacial film intensified, diverting the flow outside the sample. Consequently, precise control of both the sample dimensions and the dispensed fluid volume was necessary to achieve repeatable flow confinement. In addition, evaporation during actuation---driven by viscous dissipation and local temperature gradients, caused non-uniform fluid loss---further complicating reproducibility.

To overcome these issues, the porous samples ($4~\mathrm{mm}\times 6~\mathrm{mm}$) were laminated between standard polyester films. Two opposing edges were sealed with roughly 1~mm of UV-cured resin to eliminate leakage through air gaps at the laminate interfaces. The laminated samples were then diced so that the sealed edges remained enclosed while the opposite sides formed inlet and outlet ports. These openings were aligned with the SAW propagation direction ($x$) to favor direct acoustic pumping through the porous medium.

\section{Conclusion}
Conventional interdigital transducers are limited by rapid SAW attenuation and poor coupling into saturated porous media, which confines transport to a small region near the source and prevents practical long-range pumping. FEUDTs, whose electrodes occupy a large active area, overcome this limitation by generating SAWs across the full aperture and coupling acoustic energy more effectively through multilayer interfaces, thereby enabling directional transport across the entire porous sample. In prewetted laminated Whatman paper with 12~$\mu$m pores, FEUDTs produced flow velocities up to 0.044~mm/s, where capillary-driven flow is effectively absent. Notably, this acoustically driven velocity is 1.1 times greater than the capillary flow rate, and approximately five orders of magnitude higher than diffusion. In polyethylene porous media with 60--100~$\mu$m pores, where the characteristic pore size more closely matches the SAW wavelength, velocities reached 0.6~mm/s at sub-watt input power. Pumping was sustained over millimeter-scale distances and through samples as thick as 3.125~mm, showing that a surface-generated acoustic field can still create useful pressure gradients deep within a porous matrix. Although viscous dissipation produces an inlet-to-outlet temperature gradient and therefore a spatial viscosity gradient, these effects do not alter the governing transport mechanism. SAW-driven acoustic streaming remains dominant, and the theoretical model reproduces the measured trends, suggesting that the pore surface roughness plays an important role via an effective slip length in determining the overall flow through the porous media in our experiment. FEUDT actuation also enabled rapid, directional transport of a small-molecule tracer through porcine dermis, demonstrating both lateral transport and through-thickness penetration in a biologically relevant matrix. The results establish practical design rules linking transducer architecture, pore geometry, and actuation conditions, and they show how passive porous materials can be converted into actively pumped transport platforms.

\medskip
\textbf{Acknowledgments} \par 
J.\ Friend is grateful to the Office of Naval Research, United States (grant 13423461), the National
Science Foundation (grant ECCS 2314118), and support from UC San Diego via the Penner Endowed Chair
of Engineering and Physics and from Washington University in St.\ Louis via the Stephen F.\ and Camilla T.\ Brauer Endowed Chair of Mechanical Engineering for support of this work. O.\ Manor is grateful for support from the Ministry of Innovation, Science $\&$ Technology of the State of Israel (grant 1001827275) and from the Israel Science Foundation (grant 520/24).
\medskip

%

\newpage
\bibliographystyle{unsrtnat}
\bibliography{references}
\end{document}